\documentclass[10pt, preprint2]{aastex}

\slugcomment{Beyond the Milky Way}

\shorttitle{Solid Pulsars}
\shortauthors{R. X. Xu}

\begin{document}

\title{A note on solid pulsars}

\author{R. X. Xu}
\affil{School of Physics, Peking University, Beijing 100871,
China}

\begin{abstract}

After a brief introduction to the necessary of {\em quark stars}
in modelling pulsars, I present a qualitative analysis of the {\em
solidification} of quark matter with low-temperature but
high-density. The reason, that a solid neutron star could {\em
not} be possible, is also given.

\end{abstract}



What's the nature of pulsars? This question seems to have been
answered: pulsars are simply {\em neutron stars} --- a kind of
boring big ``nuclei'' bound dominantly by gravity rather nuclear
force.
However, this view is criticized in both theoretical and
experimental grounds: (1) nucleons (neutron and proton) are in
fact not structureless points although they were thought to be
elementary particles in Landau's time (1930s), they (and other
hadrons) have been actually composed of {\em quarks} proposed
since 1964; (2) the neutron star model does sometimes not fit well
the recent observations of advanced X-ray facilities in space, and
there is still {\em no} convincing evidence that pulsars are just
neutron stars.
Therefore, we have to pay attention to this issue, and to
reconsider the question again, even more than 35 years later.

The neutron star ideal appeared in 1932, soon after the discovery
of neutrons by Chadwick. After the first radio pulsars was
discovered in 1967 and as the quark model of hadrons developed in
1960s, a new ideal was proposed: there may exist a kind of compact
stars to be similar to neutron stars, so-called {\em Quark stars},
that are composed mainly of quark matter, rather than neutron
matter\footnote{
{\em Quark matter} is defined as a state in which quarks and
gluons are de-confined (to be thus called also as quark-gluon
plasma), whereas {\em neutron matter} as a state of free nucleons
(neutrons and a few parts of protons).
}. %
Since quark stars can also reproduce the behaviors of pulsar-like
stars, it is argued then that both neutron and quark star models
are possible. But nature may choose only the second one of these
two if quark matter with strangeness (i.e., {\em strange quark
matter}) is absolutely stable. Quark stars with strangeness are
called to be {\em strange stars} \citep{xu03a}. Are pulsars really
neutron or strange stars? Because we can {\em not} now determine
whether strange quark matter is absolutely stable from the first
principles, we have to try to answer the question via
astrophysical experiments: observations!

Strange stars, if they are radio pulsars, were usually considered
to be crusted (i.e., there is a $\sim 10^{-5\sim -6} M_\odot$
crust around a strange quark matter core) until 1998 when
\cite{xq98} addressed that {\em bare} strange stars (i.e., strange
stars without crusts) being chosen as the interior of radio
pulsars have three advantages: 1, the spectral features; 2, the
bounding energy; and 3, the core collapse process during
supernova.
The formation of bare strange stars is possible in principle, and
their emission can generally explain various observations of
pulsar-like stars \citep{xu03b}.
Furthermore, it is suggested that pulsar-like stars could be solid
quark stars with strangeness, sometimes exposing their quark
surfaces, in order to fit the continuum thermal X-ray spectra and
to understand the free-precession behaviors \citep{xu03c}.

In this note, two selected points relevant to solid pulsars are
discussed.


1. {\em Can a solid neutron star be possible?}
The answer is {\em no}.
Free precession of isolated pulsars could be common in nature; the
gradual, long-term variation of RX J0720.4-3125 (a radio-quiet
isolated neutron star) observed in X-ray band may imply that the
star is precessing \citep{de04}, besides two wobble radio pulsars:
PSR B1828-11 \citep{sls00} and PSR B1642-03 \citep{slu01}.
However, solid pulsars are suggestive because of the difficulties
of explaining this feature in the the conventional models of fluid
neutron stars. The precession should be damped down soon due to
various dissipation processes (e.g., the vertex pinning, the
surface-rubbing of Ekman layer immediately beneath the crust), and
solid quark star model for pulsar-like stars is then proposed
\citep{xu03c}.
Recently, \cite{ld04} studied the magnetohydrodynamic coupling
between the crust and the core of a rotating neutron star, and
found that the precession of PSR B1828-11 should decay over a
human lifetime. This well-defined MHD dissipation should certainly
be important in order to test the stellar models.
A solid quark star is just a rigid body, no damping occurs, and
the solid pulsar model may survive the observational tests if the
free precession keeps the same over several tens of years.
We may expect a precession without damping even hundreds of years,
or we might be lucky enough\footnote{
The chance that we detect two precessing radio pulsars among 1500
ones with ages $\sim 10^6$ years during 30 years is just $\sim
4\times 10^{-8}$! More precessing pulsars could be found if the
detection and analysis technique develops and long-term monitoring
keeps.
}.%

A neutron star can not be in a solid state.
The nuclear liquid model is successful experimentally, the
original version of which is the nuclear Fermi-gas model. Nucleons
can be regarded as Fermi-gas due to the Pauli exclusive in the
model. A great part of neutron star matter has density to be
approximately the nuclear saturation density, and at least this
part is in a fluid state.
In this sense, a quark star is identified if one convinces that a
pulsar is in a solid state.

2. {\em Why can a solid quark star be possible?}
A quark star (or quark matter) is called to be strange if it has
strangeness. Certainly it is not necessary that quark matter is
strange \citep{cea03,shh03}, but it is conjectured that the
presence of strangeness might be energetically favorable
\citep{bodm71,wit84}. Phenomenological calculation shows that the
conjecture is very likely to be true \citep{fj84}.

For quark matter in low temperature, $T$, but high baryon density,
$n_{\rm b}$, there exists a {\em competition} between color
superconductivity and solidification, just like the case of
laboratory low-temperature physics.
Normal matter should be solidified as long as the interaction
energy between neighboring ions is at least\footnote{%
The value could be $\sim 10$ if the the matter is composted of
neutral molecular.
} %
$\sim 10^2$ times that of the kinetic thermal energy \citep{dn99}.
In this sense, high density (then a small separation between ions,
$l$) and low temperature favor a solid state. However, this
classical view is not applicable (i.e., quantum effects dominate)
when the de Broglie wavelength of ions $\lambda\sim
h/\sqrt{3mkT}>l$ ($m$ the mass of ions).
So small values of $l$ and $T$ result also in a quantum state of
the system. \underline{This is the very competition}.
We need thus {\em weak interaction} between particles and {\em low
mass} in order to obtain a quantum fluid before solidification.
This is why {\em only helium}, of all the elements, shows
superfluid phenomenon though other noble elements have similar
weak strength of interaction due to filled crusts of electrons.
The superconductivity of electrons can also be understood
accordingly.

Quark-gluon plasma is a system with elemental color interaction.
This strong interaction ({\em and} the Coulomb interaction in
system with strangeness) may be responsible for a possible
quark-clusters in quark matter with low $T$ but high $n_{\rm b}$.
Note that such a cluster differs ordinary colorless hadron very
much. The residual color interaction between the massive clusters
could conduce to a solid state of quark stars

The color superconductivity state proposed is based on the field
theories, beginning with a phenomenological Lagrangian. One theory
with this spirit, which is successful in description of nuclear
matter, is the relativistic nuclear mean field theory
\citep{glen00}, where meson fields are considered to be constant
in space-time (an approximation of classical fields).
However, the $\alpha$-decay, which is well known experimentally,
can {\em not} be understood (because of no $\alpha$-clusters) in
this theory.
Nevertheless, it is interesting whether quark-clusters can form in
quark stars according to phenomenological QCD calculations.

Anyway, further experiments (in low-energy heavy ion colliders)
might help us to answer the question in the near future --- quark
matter with low temperature but high density:
color-superconductivity {\em or} solidification?

\acknowledgments

This work is supported by National Nature Sciences Foundation of
China (10273001) and the Special Funds for Major State Basic
Research Projects of China (G2000077602).

\end{document}